\documentclass[preprint2]{aastex}

\shorttitle{Observational signatures of coronal kink instability}
\shortauthors{Botha, Arber, Srivastava}

\begin{document}

\title{Observational signatures of the coronal kink 
       instability with thermal conduction.}


\author{G. J. J. Botha}
\affil{Centre for Fusion, Space and Astrophysics, Physics Department, 
       University of Warwick, Coventry CV4 7AL, UK}
\email{G.J.J.Botha@warwick.ac.uk}


\author{T. D. Arber}
\affil{Centre for Fusion, Space and Astrophysics, Physics Department, 
       University of Warwick, Coventry CV4 7AL, UK}
\email{T.D.Arber@warwick.ac.uk}


\author{Abhishek K. Srivastava}
\affil{Aryabhatta Research Institute of Observational Sciences (ARIES), Nainital 263129, India}
\email{aks@aries.res.in}

\begin{abstract}
It is known from numerical simulations that thermal conduction 
along magnetic field lines plays an 
important role in the evolution of the kink instability in coronal loops. 
This study presents the observational signatures of the kink instability in 
long coronal loops when parallel thermal conduction is included. The 3D nonlinear 
magnetohydrodynamic equations are solved numerically to simulate the 
evolution of a coronal loop that is initially in an unstable equilibrium. 
The loop has length 80 Mm, width 8 Mm and an initial maximum twist of 
$\Phi=11.5\pi$, where $\Phi$ is a function of the radius.
The initial loop parameters are obtained from a highly twisted loop 
observed in the TRACE 171 \AA~waveband.
Synthetic observables are generated from the data. These observables include 
spatial and temporal averaging to account for the resolution and exposure times 
of TRACE images. Parallel thermal conduction reduces the maximum local temperature 
by up to an order of magnitude. 
This means that different spectral lines are formed and different 
internal loop structures are visible with or without the inclusion 
of thermal conduction.  
However, the response functions sample a broad range of temperatures.
The result is that the inclusion of parallel thermal conductivity does 
not have as large an impact on observational signatures as the order 
of magnitude reduction in the maximum temperature would suggest;  
the net effect is a blurring of internal features of the 
loop structure. 
\end{abstract}



\section{Introduction}

The magnetohydrodynamic (MHD) kink instability in cylindrical geometry 
serves as a first approximation for this instability in coronal loops. 
Its evolution has been studied numerically for more than a decade  
\citep{MikicEA90,BatyHeyvaerts96,LionelloEA98,ArberEA99}. 
Previous work has studied the onset of the kink instability by 
rotating a flux tube's footpoints until the twist increased beyond the 
critical value \citep{GalsgaardNordlund97,GerrardEA02}. 
Curvature effects have also been incorporated into the model by studying a
kink-unstable flux tube, curved in a half-torus with its footpoints 
anchored in the same plane \citep{TorokEA04,TorokKliem05,GerrardEA04}. 

In another series of studies, the ideal MHD kink instability has been evoked as 
a trigger for reconnection occurring in coronal loops 
\citep{BrowningVdLinden03,BrowningEA2008,HoodEA09,BarefordEA10}. 
A cylinder is initialised with a twisted magnetic field profile that is 
unstable, and is then perturbed to evolve into the kink instability.   
\cite{BothaEA11} have shown that by including thermal conduction along 
magnetic field lines in these models, the maximum temperature obtained 
during reconnection events is lowered by an order of magnitude, activating 
different spectral lines than when no parallel thermal conduction is included.  

In this paper the observational signatures of the MHD kink instability, 
with thermal conduction included, are presented and compared with a simulation 
without thermal conduction. A kink-unstable cylinder 
is evolved and the numerical results are compared to observations of a 
twisted coronal loop \citep{SrivEA10}. In order to facilitate the comparison 
between the observed and forward modelled loops, the initial conditions 
are obtained from the observations by \cite{SrivEA10}. The numerical  
results are filtered through the TRACE temperature response 
function before synthetic images in the 171 \AA~band are presented of 
spatially and temporally averaged line of sight intensity measurements, 
similar to \cite{HaynesArber07}.  

The paper is divided into two main parts. In the first part the model 
is described, which includes a short description of the physical model, 
the initial conditions of the numerical simulation and how the synthetic 
images were obtained from the numerical data. In the second part the 
numerical results are discussed and the synthetic intensity images are 
generated from the data. 


\section{Model}

The coronal loop is initialised as a straight twisted cylinder in a 
uniform background temperature and density. 

\subsection{Physical model}

Nonlinear three-dimensional simulations are performed using the MHD 
Lagrangian-remap code, Lare3d, as described by \cite{ArberEA01}.
It solves the resistive MHD equations for a fully ionised plasma, with a heat 
flux included in the energy equation. A full description of the physical model 
is given in \cite{BothaEA11}. 

Thermal conduction is included along the 
magnetic field in the form of the classical \cite{SpitzerHarm53}, or 
Braginskii, conductivity with $\log\Lambda=18.4$. This corresponds to 
the standard thermal conductivity parallel to the magnetic field of 
$\kappa_\parallel =10^{-11}\,T^{5/2}$ W m$^{-1}$ K$^{-1}$  
\citep{Priest00}. 
Radiative losses do not play an important role and are not included in 
the model. For the coronal values used in the model, 
the radiative cooling time is of order hours. During the simulations 
reconnection and heating occur within short lengths along the loop, typically 
one tenth of the loop length \citep{BothaEA11}. When the conductive cooling 
time is calculated for this length, it is in the order of minutes. 

The code contains an artificial resistivity that is activated only when 
and where the current exceeds a critical value. It is of the form 
\begin{equation}
\eta = \left\{ \begin{array}{ll}
                 \eta_0, & \quad |j|\geq j_c, \\
                  0,      & \quad|j|< j_c,
               \end{array} 
        \right.
\end{equation}
where $\eta_0$ is the anomalous resistivity and $j_c=2$ mA is the critical 
current. $\eta_0$ is activated as soon as the kink instability occurs 
and then it stays active for the duration of the simulation 
\citep{BothaEA11}.

\subsection{Initialisation data}

The data used to obtain the initial conditions of the simulated coronal 
loop is obtained from observations of a highly twisted loop \citep{SrivEA10}. 
The loop was situated in AR NOAA 10960 and was observed on 2007 June 4 
between 04:43 UT and 04:52 UT. SoHO/MDI, Hinode/SOT G-band (4504 \AA) 
and Hinode/SOT Ca II H (3968 \AA) were used respectively to observe the 
photospheric and chromospheric parts of the active region and the 
associated highly twisted loop, while TRACE 171 \AA~was used to observe 
its coronal part. 
Figure \ref{fig:plot1} shows a closeup of approximately half of the 
observed loop system in TRACE 171 A, 6 minutes 
after the activation of helical twist in the loop
during the flaring process. 

From the observations it is estimated that the loop length is 
$\sim$80 Mm and its radius in the corona $\sim$4 Mm. 
This aspect ratio of 1:10 is within the observed range for coronal loops
\citep{Klim00,WatKlim00,AshBoe11}. The width of 8 Mm is larger than 
the usually observed coronal loop width of 2-4 Mm. Given that active 
region coronal loops change between 40\% and 70\% in diameter between 
their base in the upper transition region or lower corona and their apex 
\citep{BrooksEA07}, we assume that the loop radius at its footpoints in 
the upper transition region is 2 Mm. One of the footpoints of the active 
region loop was above a positive polarity sunspot \citep{SrivEA10} and 
it is estimated that the lower bound of the magnetic field strength 
at this location was approximately 470 G. 
The minimum average strength of the magnetic field is estimated by 
assuming the homogeneous distribution of measured magnetic
fluxes above the selected region over the positive polarity sunspot.
However, the strength of the magnetic field may change
at various locations above it between its minimum and maximum 
values. From this photospheric value 
the chromospheric magnetic field strength is calculated. 
\cite{PetPat09} found that the photospheric field is mostly vertical 
while the chromospheric field has no preferred direction. Since we 
are considering a magnetic loop, the assumption is made that most of 
the photospheric magnetic field goes to the chromosphere. However, this 
would be an upper limit. \cite{KozSom09} looked at 36 sunspots and found 
that from a height of 100 km to 1850 km the magnetic field strength 
diminishes between 0.1 and 0.7 G km$^{-1}$, with the mean field gradient 
$0.35\pm$0.05 G km$^{-1}$. Using a magnetic field gradient of 0.2 G km$^{-1}$ 
and the field strength at the photosphere, the magnetic field strength 
at a height of 1850 km is calculated to be 80 G. If we assume that the 
field strength changes little between the upper chromosphere and the 
upper transition region, then the flux through the loop footpoint 
(radius 2 Mm) is $10^{19}$ Mx. With the assumption of flux conservation 
along the loop, the magnetic field strength at the loop apex 
(radius 4 Mm) is 20 G. This lead us to initialise the loop with a 
maximum field strength of 20 G inside the loop, while the background 
outside the loop contains a uniform field of 15 G that is parallel to 
the cylindrical axis. 

In the numerical simulations the coronal loop is initialised as a uniform cylinder 
in force-free equilibrium and unstable to an ideal MHD kink instability \citep{HoodEA09}. 
The axial twist is given by 
\begin{equation}
\Phi = \frac{LB_\theta}{rB_z} \quad\mbox{with}\quad \max(\Phi)=11.5\pi,
\label{eq:twist}
\end{equation}
where $L$ is the loop length and $r$ is its radius. 
The axial magnetic field is given by $B_z$ and the azimuthal field by 
$B_\theta$. Both $B_\theta$ and $B_z$ are functions of $r$. The 
maximum twist is at position $r=1$ Mm, with zero twist on the loop axis 
and at its edge. The radial profile of the twist is presented in Figure 2(b) 
in \cite{HoodEA09}.
The maximum number of twist (six full turns) is chosen from the observational 
analysis of the highly twisted loop system shown in Figure \ref{fig:plot1}, where 
three full turns are clearly visible. The total of six turns is estimated by 
extrapolating to the invisible part of the loop by assuming a symmetric loop 
shape. Max$(\Phi)$ exceeds the stability threshold so that the loop is 
kink unstable. The critical value for twist in a cylinder was found numerically 
to be $4.8\pi$ \citep{MikicEA90}, while linear theory predicts it to be 
$2.5\pi$ \citep{GerrardEA02}. 
\cite{GalsgaardNordlund97} found that the critical twist angle is a function of 
the loop diameter, field strength and magnetic resistivity in the model. 
A full description of the magnetic field 
structure used in the simulations is given in \cite{BothaEA11}. 

Gravity is absent in the simulations. 
The initial temperature and mass density are chosen to be uniform and constant. 
The mass density has a value of $1.67\times10^{-12}$ kg m$^{-3}$ \citep{YoungEA09} 
and the temperature 0.125 MK. This temperature was chosen to make the evolution 
of the kink instability visible for the TRACE satellite.  
During the evolution of the kink instability the temperature increases locally 
where reconnection occurs \citep{BothaEA11}. Through trial and error the initial 
temperature was chosen so that these high temperatures will be visible using the 
TRACE 171~\AA~temperature response function.

The evolution of the coronal loop is studied in Cartesian geometry 
(Figure \ref{fig:cartoon}). 
The boundaries in the $(x,y)$ plane perpendicular to the loop axis are at 
$\pm 8$ Mm and reflective. Thus, there are 4 Mm between the loop's edge and 
the outer boundaries. This distance proved to be adequate so that 
no feedback from the boundaries influences the numerical results. 
Along the loop axis the boundaries are at $\pm 40$ Mm with velocities held at 
zero and the temperature fixed at the initial background value but allowing 
temperature gradients. Hence, a heat flux across the ends of the loops exists. 
The grid resolution in $(x,y,z)$ is given by $128\times 128\times 256$.

\subsection{Image generation}

The images generated from the numerical simulations were obtained by using the 
temperature response functions of TRACE \citep{SchrijverEA99} as well as those 
of SDO/AIA \citep{AshBoe11}. Details of the line contributions for the TRACE 
channels are in \cite{HandyEA99} and for the AIA channels in \cite{ODwyerEA10}. 
The emission is calculated at every node of the numerical grid and then integrated 
along the $y$ direction perpendicular to the axis of the loop, as indicated in 
Figure \ref{fig:cartoon}. The line of sight integral is given by  
\begin{equation}
I = \int^{+L_y}_{-L_y} g(T)\rho^2 dy
\label{eq:line}
\end{equation}
where $I$ is the measured intensity, $g(T)$ the temperature response function of 
the respective instruments and $\rho$ the mass density. Figure \ref{fig:responses} 
presents the temperature response functions of TRACE and AIA for 171 \AA. This 
integration produces intensity images in the $(x,z)$ plane, which are then 
integrated over time, with the time interval determined by the exposure time of 
the instrument. For TRACE this is taken to be 31.9 s and for SDO/AIA the time 
interval is 2.9 s. Finally, the time integrated image is degraded by spatially 
averaging over squares of $0.375\times 0.375$ Mm$^2$ to compensate for the pixel 
resolution of both TRACE and SDO/AIA. The process is illustrated in Figure 
\ref{fig:compare}, using the TRACE 171 \AA~temperature response function as given 
in Figure \ref{fig:responses}.


\section{Discussion}
\label{sec:disc}

Two simulated data sets are presented: with and without thermal 
conduction parallel to the magnetic field lines. 
All other parameters and initial conditions are kept identical 
between the two runs. The physical 
consequences due to the inclusion of parallel thermal conduction 
are discussed by \cite{BothaEA11}. In this paper the line of sight 
intensity contours are presented as observed through TRACE 
response functions. As such, only physical processes that 
are needed to explain the observables are included. The density and 
temperature from both data sets are used with TRACE response 
functions to generate intensity contours that can be compared with 
observational data. 

After initialisation, the kink unstable loop evolves through a linear 
phase that lasts for 300 s, as is evident from Figure \ref{fig:kink}.  
During the nonlinear phase the kink instability drives magnetic field 
into a current sheet, where reconnection occurs and the temperature 
reaches a maximum due to the energy released. Subsequently, thermal 
conduction along magnetic field lines transports heat along the 
magnetic field lines away from the points where the plasma was 
heated by reconnection. The time scale for the evolution of the kink 
instability is the same with and without thermal conduction 
(Figures \ref{fig:kink} and \ref{fig:kink3}). In the case without 
thermal conduction the heat due to reconnection is not conducted 
along magnetic field lines and the temperature maximum is higher. 
This can be seen in Figure \ref{fig:kink} where the maximum temperature 
reached without parallel thermal conduction is 9 MK, compared to a 
maximum of 3 MK with thermal conduction. 
The physical processes during the kink instability, with and without 
thermal conduction, are discussed in more detail by \cite{BothaEA11}. 

The emission, as observed in the TRACE 171 \AA~band, is presented in 
Figures \ref{fig:TRACEobserve} and \ref{fig:TRACEnoK} for the cases 
with and without parallel thermal conduction. 
Figure \ref{fig:responses} shows 
that the temperature response function lies within the range determined 
by the minimum and maximum temperatures from the simulation with 
thermal conduction (Figures \ref{fig:kink} and \ref{fig:kink3}). 
The higher maximum temperature obtained without thermal conduction 
(Figure \ref{fig:kink}) causes the temperature response function to 
sample different aspects of the evolution of the kink instability. 
However, Figure \ref{fig:kink3} show that the average loop 
temperatures with and without thermal conduction are 
comparable, making the intensity images in Figures 
\ref{fig:TRACEobserve} and \ref{fig:TRACEnoK} not radically 
different. This is due to the fact that heating occurs only at 
small localised areas, from which heat is then conducted along 
magnetic field lines. 
Figures \ref{fig:TRACEobserve} and \ref{fig:TRACEnoK} show that the 
kink instability causes the same structures to form in both cases. 
Thermal conductivity causes the temperature to spread along magnetic 
field lines, resulting in images of which the features are less 
defined when compared with the images generated from data without 
thermal conductivity. 

Irrespective of the inclusion or exclusion of thermal conductivity, 
the emission images show that the footpoints of the loops increase 
their emission during the nonlinear phase of the kink instability. 
The line of sight integral (\ref{eq:line}) is determined by the 
temperature response function as well as the mass density. Figure 
\ref{fig:tempdens} gives the temperature and mass density profiles 
along the central axis towards the end of the simulation for 
the case when thermal conduction is included. 
Note that the central axis experiences some heating (it has higher 
temperatures than the average temperatures in Figure \ref{fig:kink3}) 
but the locations with maximum temperature are not on it (as it has 
lower temperatures than the maxima in Figure \ref{fig:kink}).
Figure \ref{fig:tempdens} shows that the 
enhanced emission is due to a density increase at the footpoints 
-- and not footpoint heating. Plasma flows from the middle of the 
numerical domain, where current sheets form and reconnection 
occurs, driven by MHD ponderomotive forces generated during the 
kink instability. Footpoint brightening due to compression was also 
observed in the coronal loop simulations of \citet{HaynesArber07}. 

One factor that determines the onset of the nonlinear phase of the kink instability is 
the initial twist (\ref{eq:twist}) in the coronal loop. A smaller value of $\max(\Phi)$ 
increases the duration of the linear phase. However, it was found that once the nonlinear 
phase is reached, the formation of the current sheet, the reconnection and the thermal 
aftermath have the same duration as long as the initial twist exceeds the stability 
threshold. 

The first image in Figure \ref{fig:TRACEobserve} is sampled at 
290.0 s and shows the twisted structure towards the end of the linear phase of the kink 
instability. In the second image (at 321.9 s) the nonlinear phase has formed a current 
sheet and in the third image (at 353.8 s) thermal conduction has transported heat from 
the reconnection site along the magnetic field lines. From 321.9 s the evolution of 
the internal loop structure along its length is observed for 4 minutes until the end 
of the numerical run at 577.0 s. As reconnection occurs within the loop, magnetic 
field lines straighten out along the length of the loop \citep{HaynesArber07}. 
At the same time heat is transported along magnetic field lines away from reconnection 
sites, so that the loop cools and its internal structure becomes less defined
\citep{BothaEA11}. 

The differences in intensity images with and without thermal 
conduction are clear when the intensity images at 449.5 s and later times in 
Figures \ref{fig:TRACEobserve} and \ref{fig:TRACEnoK} are compared. 
Despite the order of magnitude reduction in the maximum temperature 
when thermal conduction is included (Figure \ref{fig:kink}), similar 
looking internal loop structures are 
visible in the intensity images with and without thermal conduction. 
The response functions capture a range of temperatures (Figure \ref{fig:responses}) 
and because the average temperatures in the coronal loops are of similar values 
(Figure \ref{fig:kink3}), the observational signatures are less sensitive to the 
inclusion of parallel thermal conduction. 
The main effect due to the inclusion of 
thermal conduction is the blurring of the internal features of the coronal loop, 
which is due to the efficient conduction of heat along magnetic field lines. 
In the simulation with thermal conduction the maximum temperature of the magnetic 
structures is closer to the average loop temperature (Figure \ref{fig:kink3}) than 
the maximum temperature obtained without thermal conduction.
Consequently the internal loop structure are not as clearly delineated in the 
emission intensity as when no thermal conduction is present, which can be 
seen when comparing Figures \ref{fig:TRACEobserve} and \ref{fig:TRACEnoK}. 

In Figures \ref{fig:TRACEobserve} and \ref{fig:TRACEnoK} it appears that 
the amount of twist towards the end of the linear phase (at 290 s) is 
approximately three turns -- instead of the six turns dictated by 
$\max(\Phi)$ in (\ref{eq:twist}). The heating caused by the kink 
instability is highly localised and only magnetic field lines that 
pass through the heated area are visible. The physical processes 
associated with this are discussed by \cite{BothaEA11}. As the twist 
is a function of the radius, one should not expect the line of sight 
integral (\ref{eq:line}) to capture the heating that occurs at 
$\max(\Phi)$.

When comparing the numerical results with observations, it is more realistic 
to compare the results obtained when thermal conduction is included in the model. 
Figure \ref{fig:plot1} shows a loop segment of the observations from \cite{SrivEA10}
at a time in its evolution when the twisted threads are most visible in the observed 
images. The simulation images in Figure \ref{fig:plot1} shows a similar fine internal 
structure, but it is clear that a perfect visual or quantitative match is not possible.  
As an example of this, Figure \ref{fig:TRACEobserve} at time 353.8 s or 385.7 s shows 
the twist to be one full turn over 40 Mm, while the observed image shows a full turn 
to be over 20 Mm. 

In this paper the results were obtained from only one initial equilibrium state. 
It is possible that other initial conditions may lead to end results that 
correspond closer with the observations by \citet{SrivEA10}. This needs further 
investigation. 
We also ignored curvature and gravity, both of which may be important 
during the evolution of the kink instability. Assuming the shape of the loop to 
be a perfect half-circle, the top of the loop will be 25 Mm above the transition 
region. The average and maximum temperatures inside the loop during the evolution 
of the kink instability are approximately 0.3 MK and 2 MK. These temperatures give 
gravitational scale heights of approximately 20 Mm and 120 Mm. 
Thus, density stratification is important and gravity cannot be ignored in the 
forward modelling of the kink instability for this loop. The inclusion of both 
curvature and gravity should be investigated further.

The observations by \citet{SrivEA10} show that after the heating event, the loop 
structure cools down within five minutes to photospheric temperatures. The duration 
of the simulations presented here, shows 4.5 minutes of the nonlinear evolution of the 
kink instability. During this time the average temperature of the loop increases 
(Figure \ref{fig:kink3}) while the maximum temperature (with thermal conduction 
included) stays approximately the same (Figure \ref{fig:kink}). 
The reason for this is that as the current sheets evolve, multiple reconnection 
acts as a continuous heating source to the plasma inside the loop, with the location 
of the heating changing as the current sheets change \citep{HoodEA09}.

The temperature response function for the SDO/AIA 171 \AA~band is presented in 
Figure \ref{fig:responses}. The emissions in 171 \AA~for SDO/AIA and TRACE are 
essentially the same, because their temperature response functions lie so close 
together. The only difference between the two observational platforms is the time 
resolution. With the TRACE exposure time 11 times longer than that of SDO/AIA, the 
images from SDO/AIA are sharper.  

In addition to the figures presented in this paper, the response functions from 
TRACE 195 \AA~and 284 \AA~as well as those for the coronal bandwidths of SDO/AIA 
were used in calculating the line of sight integral (\ref{eq:line}).
In all cases a broad range of temperatures are sampled from the simulated data,
with a large part of the temperature range common to all the response functions. 
As a result, the differences between the line intensity plots were minor; 
the same basic internal loop structure was observed with no new information to 
be gained from them.


\section{Conclusion}
\label{sec:concl}

The evolution of a coronal loop is studied, solving the resistive MHD equations for a 
fully ionised plasma and with parallel thermal conduction included in the model. The loop is 
initialised as a straight cylinder with a twist above the stability threshold, 
which leads to the kink instability. Line of sight emission intensities were 
calculated of the simulation data, using the temperature response functions 
from TRACE. 

\begin{figure}
\centerline{
\includegraphics[width=12cm,angle=90]{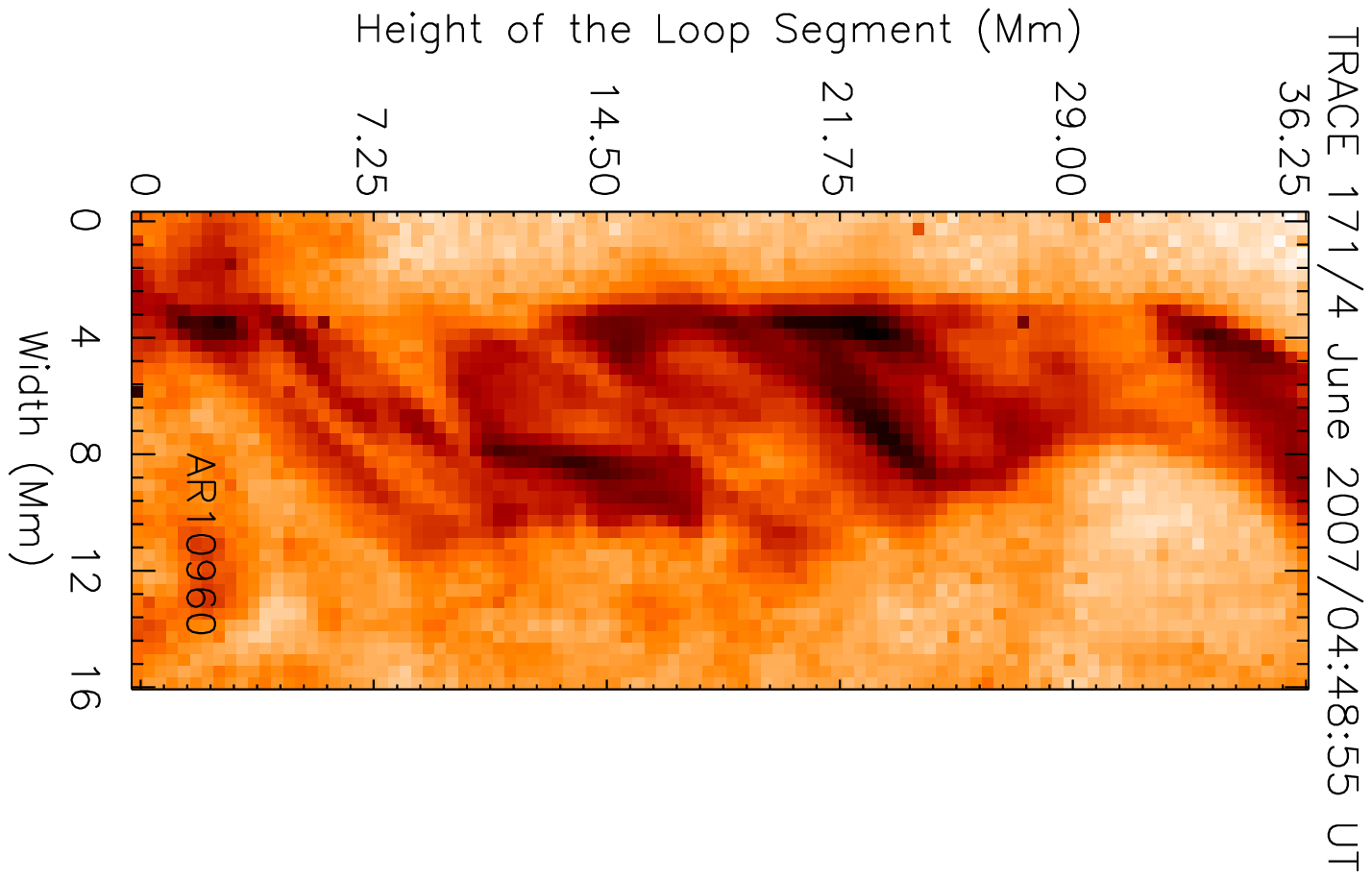}
}
\caption{The observed loop system used 
         as source for the initialisation of the simulations. Half  
         the loop length is shown with the helicity of right handed 
         twist clearly visible. A time evolution of the complete loop 
         system is presented by Figure 5 in \cite{SrivEA10}. The image 
         is in reverse colour.
         }
\label{fig:plot1}
\end{figure}

\begin{figure}
\centerline{
\includegraphics[width=4.5cm]{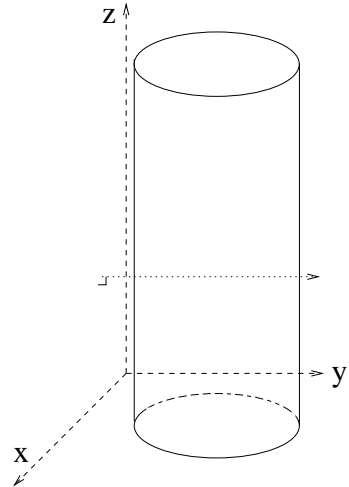}
}
\caption{Cartoon showing the orientation of the cylindrical loop in the 
         Cartesian numerical domain. The dotted line indicates the 
         integration path of the line of sight integral (\ref{eq:line}) 
         along the $y$ direction and perpendicular to the $(x,z)$ plane.
         }
\label{fig:cartoon}
\end{figure}

\begin{figure}
\plotone{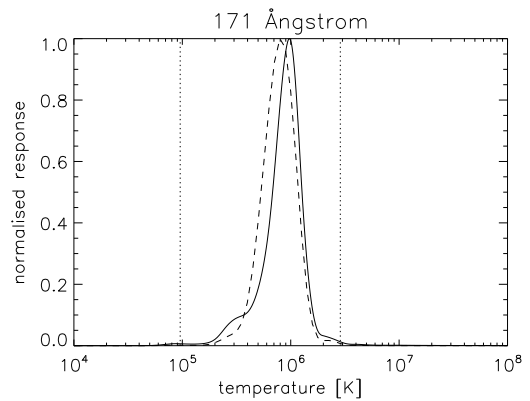}
\caption{Normalised temperature response functions for TRACE 171 \AA~(solid line) and 
         AIA 171 \AA~(broken line). 
         The two vertical dotted lines are respectively the minimum and maximum 
         temperatures during the simulation with parallel thermal conduction included, 
         as measured in Figures \ref{fig:kink} and \ref{fig:kink3}.
         }
\label{fig:responses}
\end{figure}

\begin{figure*}
\centerline{
\includegraphics[width=10cm]{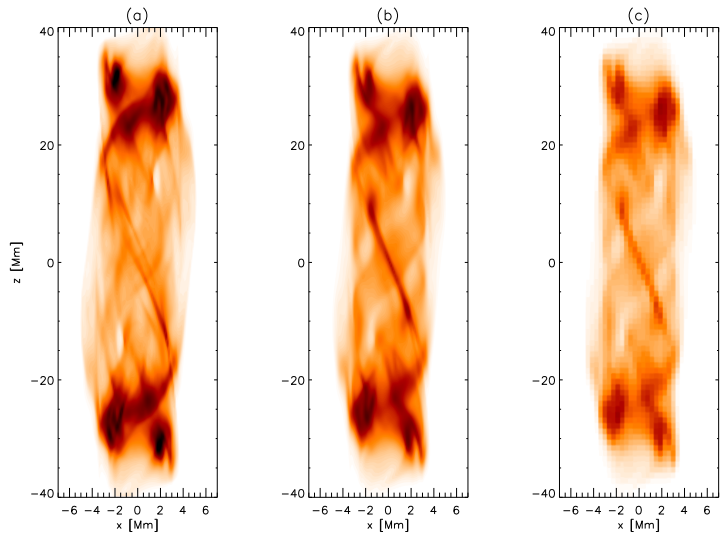} }
\caption{Intensity from the line-of-sight integration (\ref{eq:line}), using 
         the simulation data as filtered through the TRACE 171 \AA~temperature 
         response function in Figure \ref{fig:responses}:
         (a) the simulation data at 417.6 s; 
         (b) the time averaged data set;  
         (c) the time and spatially averaged data set. 
         The time exposure is 31.9 s and the spatial resolution is 0.375 Mm 
         per pixel. 
         The images are in reverse colour, with white representing the lowest 
         value on the scale. The minimum and the maximum values of the 
         reverse colour scale are the same as those used in Figure 
         \ref{fig:TRACEobserve}. These data sets were taken from the simulation 
         with parallel thermal conduction. 
         }
\label{fig:compare}
\end{figure*}

The simulations were initialised with physical parameters extracted from the 
observations of a coronal loop shown in Figure \ref{fig:plot1} \citep{SrivEA10}. 
Figure 5 in \cite{SrivEA10} shows the time evolution of a helically twisted 
structure of a flaring loop in TRACE 171 \AA~that is observed for 4 minutes. 
The initial magnetic field structure used in the simulations is a kink-unstable 
force-free equilibrium where the twist varies with radius \citep{HoodEA09}. 
This field structure was used in previous studies of the kink 
instability (see \cite{BothaEA11} and references therein) and although its 
maximum twist is similar to the global estimated twist from the observations by  
\cite{SrivEA10}, there is no guarantee that it is similar to the magnetic field 
of the observed loop. 
Throughout the evolution of the kink instability, the internal structure of the 
simulated loop is shown in the generated images (Figure \ref{fig:TRACEobserve}). 
This structure evolves into a simpler configuration as the kink instability causes 
multiple reconnection events, which have the effect of straightening the internal 
magnetic field. In contrast, the observations by \cite{SrivEA10} show a field 
structure that stays largely intact for the duration of the observations. 

Footpoint brightening due to compression of the plasma is observed in the simulation 
results, but is absent from the observations by \cite{SrivEA10}. In the simulation 
this is due to the impenetrable boundaries of the numerical domain, which cause 
the plasma density to increase at the top and bottom boundaries of the loop axis
(Figure \ref{fig:tempdens}). 
On the Sun plasma moves through the transition region to the chromosphere and photosphere. 
It may be that footpoint brightening occurs in these lower regions, but the plasma 
is so cool that most likely it will not be captured by the response functions for 
coronal temperatures in spite of collective footpoint heating, although  
\cite{SrivEA10} have observed brightpoints at coronal temperatures that may be evidence 
of localised footpoint heating.

\begin{figure}
\plotone{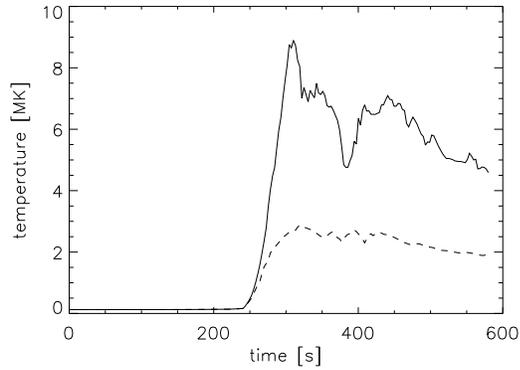}
\caption{Maximum temperatures during the evolution of the kink instability, for the 
         two numerical runs with thermal conductivity included (broken line) and 
         without thermal conduction (solid line). 
         }
\label{fig:kink}
\end{figure}

\begin{figure}
\plotone{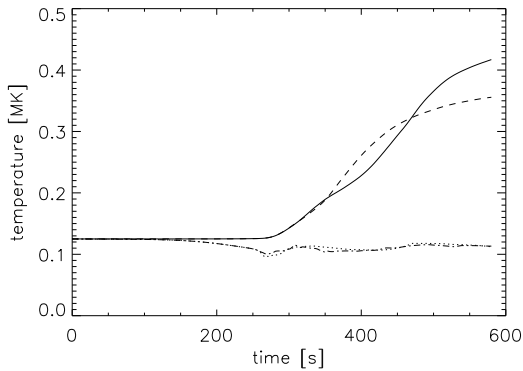}
\caption{Temperatures during the evolution of the kink instability for the two 
         numerical runs with and without thermal conductivity. The mean 
         temperature is represented by a solid line for the simulation without 
         thermal conduction and a broken line with thermal conductivity, 
         similar to Figure \ref{fig:kink}. The mean temperature is calculated 
         by averaging over the entire data cube. 
         The minimum temperatures are also included, 
         with the dash-dot-dashed line the simulation without thermal conduction and 
         the dotted line with thermal conductivity. 
         }
\label{fig:kink3}
\end{figure}

\begin{figure*}
\includegraphics[width=16.5cm]{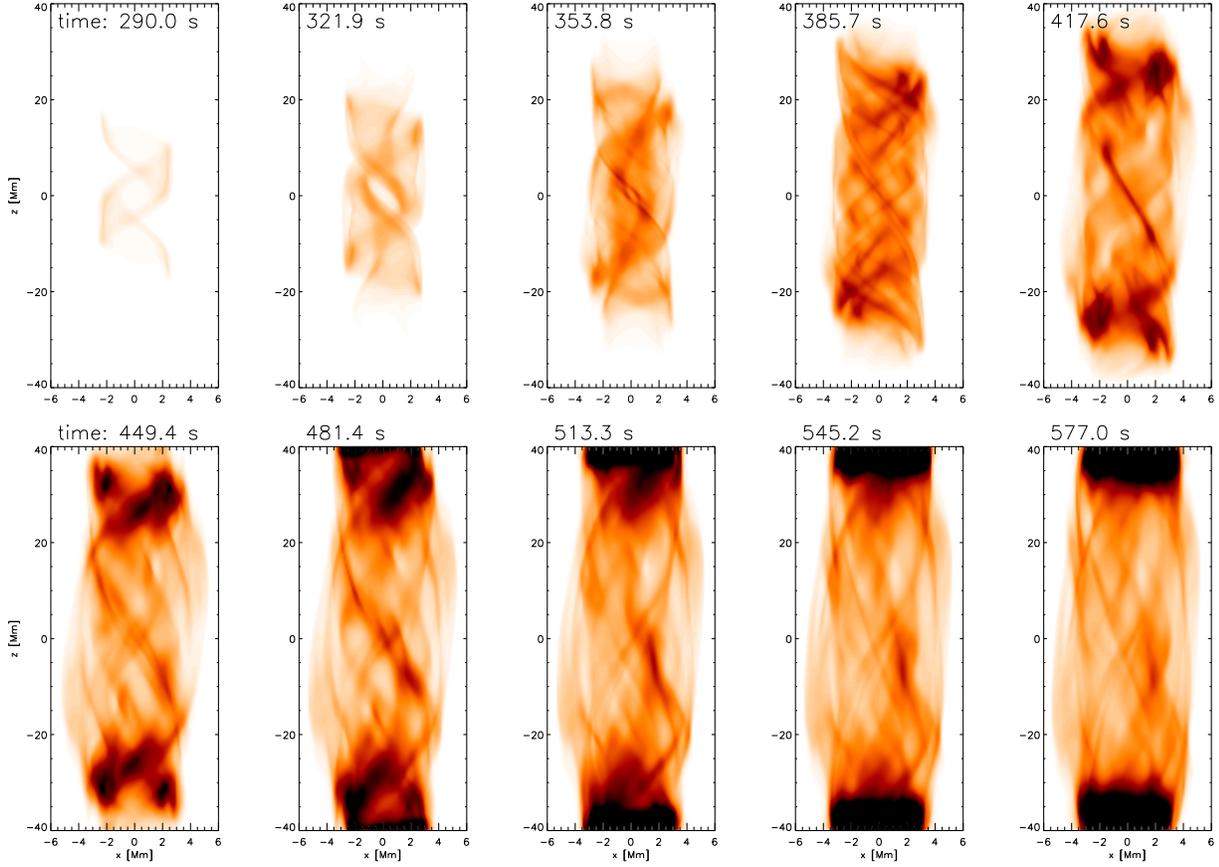}
\caption{Simulated intensities as would be seen through the TRACE 171 \AA~filter
         of the kink and its aftermath, with thermal 
         conduction included in the calculation. 
         The first image is taken at 290.0 s and the cadence is 31.9 s.
         The time exposure, spatial resolution and light intensity scale are the 
         same as in Figure \ref{fig:compare}. The reverse colour table is such that 
         the minimum intensity (white) is chosen so that the lowest 10\% of the 
         simulation values at time 290 s are eliminated from these plots. All the 
         exposures use the same reverse colour scale. 
         }
\label{fig:TRACEobserve}
\end{figure*}

\begin{figure*}
\includegraphics[width=16.5cm]{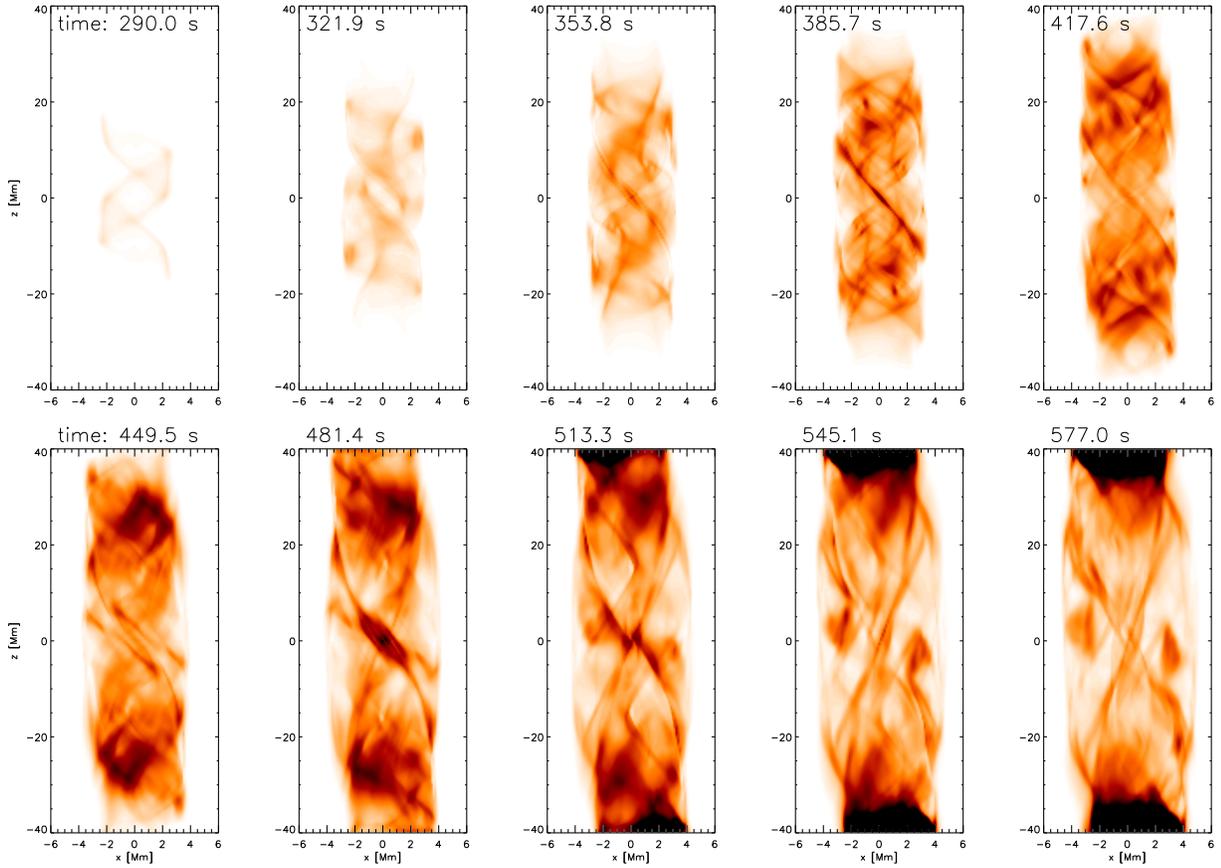}
\caption{Simulated intensities as would be seen through the TRACE 171 \AA~filter
         of the kink and its aftermath, without thermal 
         conduction. The exposure time, spatial resolution and light intensity scale 
         are the same as in Figure \ref {fig:compare}, with the minimum intensity 
         (white) and the maximum (black) of the reverse colour table having the 
         same values as in Figure \ref{fig:TRACEobserve}. The images are taken at 
         the same times as those in Figure \ref{fig:TRACEobserve}. 
         }
\label{fig:TRACEnoK}
\end{figure*}

\begin{figure}
\plotone{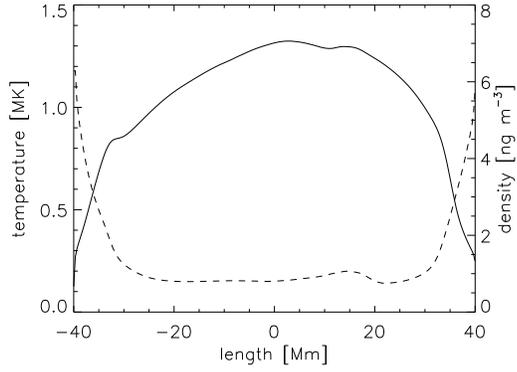}
\caption{Temperature and mass density profiles along the length of the loop on 
         its central axis at time 579.98 s, obtained from the simulation with 
         thermal conduction. 
         The solid line is the temperature and the dashed line the density. 
         Initialisation is with a constant temperature of 0.125 MK and a 
         constant density of 1.67 ng m$^{-3}$.
         }
\label{fig:tempdens}
\end{figure}

Thermal conduction in the model conducts heat along magnetic field lines.  
The kink instability heats up the plasma where the current sheet causes reconnection.  
With thermal conduction included, this heat is transported along the magnetic field lines 
away from the reconnection sites. As a result, simulations 
without thermal conduction reach maximum temperatures of up to an order of 
magnitude larger than when thermal conduction is present (Figure \ref{fig:kink}
and \cite{BothaEA11}). 
In contrast, when the line of sight integral (\ref{eq:line}) is calculated, the 
response functions of TRACE capture a broad range of temperatures. The average 
temperatures with and without thermal conduction are similar (Figure \ref{fig:kink3}) 
and as a result the observational difference due to the inclusion of thermal 
conduction is much less, as can be seen when comparing Figures \ref{fig:TRACEobserve} 
and \ref{fig:TRACEnoK}.  

This paper considers the observational effects of including parallel thermal 
conduction into the model. Moving towards more realistic coronal loop 
simulations, a temperature profile and gravity need to be included, 
the latter of which adds density stratification to the system. Curvature is 
another part that is missing from the present study, which should be 
included at a later stage. 


\acknowledgments

AKS thanks Shobhna Srivastava for patient encouragements.



\clearpage


\clearpage

\clearpage

\clearpage

\clearpage

\end{document}